\documentclass[11pt,a4paper]{article}

\usepackage{natbib}
\usepackage{hyperref}
\usepackage{amsmath}
\usepackage{amssymb}
\usepackage{graphicx}
\usepackage{subfig}
\usepackage{color}
\usepackage{dsfont}
\usepackage{bm}

\title{Spatial Interpolation of Extreme Values}
\date{\today}
\author{B. D. Youngman\footnote{School of Mathematics and Statistics, University of Sheffield, Hicks Building, Hounsfield Road, Sheffield, UK, S3 7RH; b.youngman@sheffield.ac.uk}}

\definecolor{grey}{gray}{0.5}

\begin{document}

\maketitle

\section{Introduction}

When modelling extremes of environmental phenomena often we wish to understand their behaviour over a region, in particular dependence between the extremes at different locations. This is typically hindered by two things: first that extreme events are by definition rare, and second a lack of locations where data have been gathered. Many situations exist in which understanding dependence between extremes is important, especially for environmental phenomena. For example here interest lies in estimating extreme rainfall. If areal estimates can be produced then rainfall amounts accumulated over a river's catchment could be understood and in turn this could lead to estimates of susceptibility to flooding which are vital for the insurance industry. Ship building is another area in which an understanding of dependence between extremes is important because the level of punishment experienced by a ship on a given journey will be affected by the level of dependence between sea waves at different locations. 

The aim here is to produce estimates of extreme rainfall for a large region of the UK where conventional time series data exist but only from rain gauges at a small number of locations. To overcome the spatial sparsity of the data, they will be supplemented with simulator output---from a regional climate model for example---in order to benefit from the simulator's richer spatial provision, which can typically be specified. As a result we hope to improve estimates of extreme rainfall over the region under study. While we focus on estimating extreme rainfall, there are many different simulators for many different phenomena, and the robust approach that we take can be extended to many other applications. For example, we consider only conventional data from rain gauges, though in the ship building example wave height data may come from buoys, oil-rig-mounted equipment or even satellites, all of which may be spatially sparse, but all may be supplemented with simulator output to bring more accurate spatial estimation of extreme wave heights.

The remainder of this paper is as follows. In \S\ref{spat} we outline univariate results for modelling extremes, introduce extensions to the methodology to incorporate spatial dependence and conclude by showing how model parameters may be estimated. In \S\ref{interp} we establish a link with and describe previous approaches to downscaling extremes, introduce notation, outline our proposed method for spatial interpolation of extremes and finally extend the spatial model for extremes to incorporate this. In \S\ref{check} we describe a variety of checks to assess the fit of the model. Then in \S\ref{rain} we analyse extreme rainfall for a central region of the UK using the model. Finally in \S\ref{discuss} we summarise the work presented.

\section{Spatial modelling of extremes} \label{spat}

\subsection{Univariate background} \label{uni}

This section primarily describes the underlying class of spatial extremal models that will be used in subsequent modelling of extreme rainfall, beginning with the original asymptotic extremal theory on which the model is based. Consider a strictly stationary sequence $\{Z_i\}, i=1, \ldots, n,$ and define $M_n=\max_{i=1, \ldots, n}Z_i$. If constants $a_n>0$, $b_n$ exist such that as $n\to \infty$ then \begin{equation} \label{gevconv} \text{pr}[a_n^{-1}(M_n-b_n) \leq z] \to G(z)\end{equation} where $G$ is a nondegenerate distribution function then $G$ is the generalised extreme value (GEV) distribution \begin{equation} \label{gevform} G(z) = \left\{ \begin{array}{ll} \exp\Big[-\Big(1+\xi\dfrac{z-\mu}{\psi}\Big)^{-1/\xi}\Big] & \text{if}~~\xi \neq 0,\\[2ex] \exp\Big[-\exp\Big(-\dfrac{z-\mu}{\psi}\Big)\Big] & \text{if}~~\xi = 0,\end{array}\right.\end{equation} defined when $\psi>0$, for $\{z\,:\,1+\xi(z-\mu)/\psi>0\}$ and where the case $\xi = 0$ results from the limit $\xi \to 0$.

Relying on the asymptotic results of equations \eqref{gevconv} and \eqref{gevform} and by assuming equation \eqref{gevconv} to be approximately true for sufficiently large $n$, a statistical model may then be formed for a sequence of data $z_1, z_2, \ldots$ by dividing it into plausibly homogeneous blocks all of size $n$ and then assuming that the resulting block maxima follow a GEV distribution. Quantiles of the GEV distribution have a more natural interpretation than its parameters themselves, and are commonly reported from an extremal analysis. Specifically if $q_p$ satisfies $G(q_p)=1-1/p$ then it is referred to as the $p$-year return level. For a stationary sequence it may be regarded as the level above which only one exceedance is expected in $p$ years. Based on equation \eqref{gevform} $q_p$ is given by  \[ q_p=\left\{\begin{array}{cl}\mu-\dfrac{\psi}{\xi}(1-y_p^{-\xi}) &\text{when}~\xi \neq 0,\\ \mu-\psi\log(y_p) &\text{when}~\xi = 0,\end{array}\right. \] where $y_p=-\log(1-p).$

\subsection{Spatial framework} \label{spat-frame}

The GEV model is now extended to a spatial context, with particular emphasis placed on modelling environmental phenomena. Assume that at each point $s$ in some region $R \subset \mathds{R}^2$ time series data for some process exist which are divided into blocks resulting in block maxima $X_t(s)$, $t=1, 2, \ldots$. To ensure approximately similar behaviour within blocks and equal block sizes, a common choice for environmental data is to use annual maxima. Here when modelling extreme rainfall we will consider annual maxima of daily rainfall data, more details of which will emerge in the later application. At each location $s$ assume that \begin{equation} [X_t(s)\,|\,\mu(s),\,\psi(s),\,\xi(s)]~\text{is}~GEV\big(\mu(s),\,\psi(s),\,\xi(s)\big) \label{data1} \end{equation} where $[\cdot]$ denotes ``distribution of''. The spatial model will adopt a hierarchical structure and relation \eqref{data1} will be referred to as its \emph{data layer}. For environmental data it is often the case that the dependence between $X_t(s)$ and $X_t(s')$, for locations $s, s' \in R$, relates to their relative locations in space, or more simply to their distance apart. We capture this through the GEV parameters by letting dependence exist between $\big(\mu(s), \psi(s), \xi(s)\big)^T$ and $\big(\mu(s'), \psi(s'), \xi(s')\big)^T$ and decay as a function of distance. Furthermore, \emph{all} spatial dependence is assumed to be characterised through the GEV parameters so that consequently $X_t(s)$ and $X_t(s')$ are \emph{conditionally independent} given their respective GEV parameters, for all pairs $s, s' \in R$, which we shall refer to as the conditional independence assumption.

As first used by \cite{casson-coles}, and in subsequent variants by \cite{fawc2}, \cite{cooley} and \cite{sang}, for example, we use a Gaussian process (GP) to characterise dependence between GEV parameters. For the present application the GP offers many benefits: the ability to be used for high-dimensional problems, ie. for data at many locations; ease of spatial interpolation using conditional Gaussian arguments; and the plausibility of the joint and marginal assumptions about variability induced on GEV parameters. First consider a GP assumption for the GEV location parameter $\mu(s)$. This forms one \emph{spatial process layer} of the hierarchical model in which \begin{equation} [\mu(s)]~\text{is}~GP\big(m(s), \sigma^2 c(~,~)\big), \label{proc1} \end{equation} for mean function $m(s)$, underlying variability $\sigma^2$ and correlation structure $c(~,~)$. Allowing $m(s)$ to depend on $s$ lets covariate effects be introduced, which is particularly attractive for environmental data. Then the belief of a decay in dependence with distance is incorporated through the correlation structure. The exponential structure offers decay in a simple and intuitive way, but here a slightly more relaxed modelling assumption is preferred and so we choose the powered exponential structure,  \begin{equation} \label{powexp} c(s,s') = \left\{\begin{array}{ll}1+\tau^2/\sigma^2 & \text{if}~||s-s'||=0,\\ \exp\{-(||s-s'||/\phi)^\delta\} & \text{otherwise},\end{array}\right.\end{equation} where $0 < \delta \leq 2$, $0 \leq \sigma, \tau$ and $ 0 < \phi$; only if $\tau > 0$ is the GP discontinuous everywhere. \cite{handspat} offer further choices of correlation structure. GPs may be assumed for $\psi(s)$ and $\xi(s)$ similarly, though it is more natural to work with $\rho(s) = \log\{\psi(s)\}$, to ensure the parameter's positivity.

\subsection{Model estimation} \label{mod-est}

To estimate model parameters for the present problem we use an adaptation of the Monte Carlo EM algorithm, introduced by \cite{weitanner}; more specific details of the algorithm related to the present problem can be found in \cite{mccull}. Here the method of parameter estimation is found to have many benefits, including not being unduly sensitive to starting values, converging reasonably quickly, depending on the level of accuracy sought, and avoiding prior specifications on parameters, such as those in $c(\,,\,)$, to which final parameter estimates can be sensitive. We outline the algorithm by considering the simplified case in which $[X_t(s)\, | \, \mu(s)]$ is $GEV\big(\mu(s),\,\psi,\,\xi\big)$ to ensure that estimation of parameters in both the data and spatial process layers is illustrated, which would not be possible if GPs were assumed for all parameters. Furthermore we adopt such a specification in the extreme rainfall application of \S\ref{rain}. The estimation procedure, however, extends readily to alternative formulations in which different combinations of GEV parameters are assumed to follow GPs.

Let $\theta_2$ denote parameters characterising the GP distribution of $\mu(s)$, so that the full parameter set is $\theta=(\theta_1,\, \theta_2)$ where $\theta_1 = (\psi, \, \xi)$. The $GEV\big(\mu(s),\psi,\,\xi\big)$ density will be denoted $f_1(\,|\,\mu(s),\,\theta_1)$ and the GP density relating to $\mu(s)$ denoted $f_2(\,|\,\theta_2)$. To achieve a maximum likelihood estimate of $\theta$, $\hat \theta$ say, based on a finite set of locations $S=\{s_1, \ldots, s_D\}$, we wish to maximise \begin{equation} \label{lhood1} \int_S \prod_{t = 1}^T \Big[ \Big\{\prod_{j=1}^D f_1\big(x_t(s_j)\,|\,\mu(s_j),\,\theta_1\big)\Big\}\,f_2\big({\bm \mu}(s)\,|\,\theta_2\big)\Big] ds,\end{equation} where ${\bm \mu}(s) = \big(\mu(s_1), \ldots, \mu(s_D)\big)^T$. The integral of equation \eqref{lhood1} is $D$-dimensional, which can either significantly hinder or even prohibit the finding of its analytical solution, in particular in spatial applications where $D$ may be large. In the standard EM approach to parameter estimation the random ${\bm \mu}(s)$ is treated as missing data giving complete data ${\bf z}=\big({\bf x},\,{\bm \mu}(s)\big)$ where ${\bf x}_t(s) = \big(x_t(s_1), \ldots, x_t(s_D)\big)$ and ${\bf x} = \big({\bf x}_1(s), \ldots, {\bf x}_T(s)\big)$. Then, taking logarithms of the likelihood in equation \eqref{lhood1}, we require parameters that maximise the expected log likelihood \begin{equation} \label{EMlhood} E\Big[\sum_{t=1}^T\Big( \sum_{j=1}^D\big[\log\big\{f_1\big(x_t(s_j)\,|\,\mu(s_j),\,\theta_1\big)\big\}\big] + \log\big\{f_2\big({\bm \mu}(s)\,|\,\theta_2\big)\big\}\Big)\,\big|\,{\bf x}\Big].\end{equation} However, the expected log-likelihood of equation \eqref{EMlhood} is again typically complex, beyond the finding of an analytical solution to its maximum. Draws from $[{\bm \mu}(s)\,|\,{\bf x}]$ can however be obtained using a Metropolis-within-Gibbs sampling procedure, and consequently a Monte Carlo estimate of the expectation in equation \eqref{EMlhood} can be achieved; efficient choice of proposals is discussed in \cite{mccull}. Let ${\bm \mu}_i(s)$, $i=1, \ldots, N$, denote draws from $[{\bm \mu}(s)\,|\,{\bf x}]$. Then for the Monte Carlo EM algorithm we require parameters that maximise \begin{equation} \label{EMlhood2} \frac{1}{N} \sum_{i=1}^N \Big\{\sum_{t=1}^T \Big(\Big[\sum_{j=1}^D \log\big\{f_1\big(x_t(s_j)\,|\,\mu_i(s),\, \theta_1\big)\big\}\Big] + \log\big\{f_2\big({\bm \mu}_i(s)\,|\,\theta_2\big)\big\}\Big)\Big\}.\end{equation} Recognising that the left- and right-hand sides of the sum in equation \eqref{EMlhood2} depend only on parameters $\theta_1$ and $\theta_2$ respectively, the sum may be divided into two sums accordingly and parameter estimates reached by maximising each sum separately.

\subsection{Uncertainty estimation} \label{sandwich}

The conditional independence assumption of \S\ref{spat-frame} implies that, given GEV parameters, annual maxima at different locations will be independent and have variance equal to their corresponding GEV distributions. For the present rainfall application, we can imagine that almost identical rainfall levels will be experienced at locations sufficiently close together, that is where we expect variability to be less than assumed GEVs. While this model misspecification will not affect parameter estimates, the Fisher information associated with the MCEM likelihood can no longer be used to give reliable estimates of parameter uncertainty. Consequently we modify the sandwich information correction, originating from works by \cite{huber}, \cite{eicker} and \cite{white}, so that it is applicable to a MCEM likelihood. 


We illustrate this modification to the sandwich information correction by considering only the data layer of the model, ie. for the parameters $\theta_1$, primarily based on the above example of potential model misspecification; however, extending this procedure to the process layer requires simple alteration. Because not all GEV parameters may be assumed to follow GPs, the case in which $[X_t(s)\,|\,\mu(s)]~\text{is}~GEV\big(\mu(s),\,\psi,\,\xi\big)$ is again considered. Let \begin{align*}\ell\big(\theta_1\,;\,x_t(s_j), \, {\bf \mu}_i(s)\big) &= \log\big\{f_1\big(x_t(s_j)\,|\,\mu_i(s_j),\, \theta_1\big)\big\} \intertext{and, with $\theta_1 = (\theta_{1, 1}, \ldots, \theta_{1, n_{\theta_1}})$, let} {\bm j}\big(\theta_1\,;\,x_t(s_j), \, \mu_i(s_j)\big) & = \nabla \ell\big(\theta_1 \, ; \, x_t(s_j),\,{\bf \mu}_i(s)\big) \intertext {with $k$th element} j_k\big(\theta_1\,;\,x_t(s_j), \, \mu_i(s_j)\big) & =\dfrac{d}{d\theta_{1, k}}\ell\big(\theta_1\,;\,x_t(s_j), \, \mu_i(s_j)\big)\end{align*} $k=1, \ldots, n_{\theta_1}.$ Then write \[J(\theta_1)=\dfrac{1}{N}\sum_{i=1}^N \sum_{j=1}^D \sum_{t=1}^T {\bm j}\big(\theta_1\,;\,x_t(s_j),\, \mu_i(s_j)\big)\, \big\{{\bm j}\big(\theta_1\,;\,x_t(s_j),\, \mu_i(s_j)\big)\big\}^T.\] Let $H(\theta_1)$ have $(l,m)$th element \[ h_{(l,m)}(\theta_1)= \dfrac{d}{d\theta_{1, l} d\theta_{1, m}}\bigg[ \dfrac{1}{N} \sum_{i = 1}^N \sum_{t=1}^T \sum_{j=1}^D \ell\big(\theta_1\,;\, x_t(s_j), \, \mu_i(s_j)\big)\bigg].\] This leads to the final estimate of the covariance matrix for $\hat \theta_1$ \[ H^{-1}(\theta_1)\,J(\theta_1)\,H^{-1}(\theta_1) \Big\vert_{\theta_1 = \hat \theta_1} \label{sandmat} \]where $\hat \theta_1$ is the estimate of $\theta_1$ that maximises the MCEM likelihood of equation \eqref{EMlhood2}.

\section{Spatial interpolation using computer simulator output} \label{interp}

\subsection{Background}

In this section we introduce a method for spatial interpolation of extremes based on supplementing field data---eg. resulting from a measurement or observation---with output from a numerical model, or \emph{computer simulator} as we shall refer to it, such as a regional climate model (RCM). Our motivation is the desire to produce predictions of extremes over an entire region that capture spatial dependence where field data are spatially sparse; consequently simulator output is also used in order to benefit from its high spatial resolution. The predictions produced will be representative of \emph{point level}, in theory allowing continuous maps for entire regions to be produced. In practice maps representing discretised regions at arbitrarily fine scales will be produced. Our motivation for spatial interpolation shares similarities with statistical downscaling, in which large-scale data are downscaled so that inferences about finer scales can be made. Due to this similarity we review a selection of its corresponding literature. The reader is referred to \cite{wilwig} and \cite{maraun} for more comprehensive reviews.

The most developed statistical downscaling methods use stochastic weather generators or transfer functions. Stochastic weather generators originate from the wet-dry day models of \cite{gabriel} in which transitions between wet and dry days have Markov structure. An extension of this by \cite{katzpar} is to assume a mixture distribution for the rainfall amount on a wet day, the parameters of which vary according to output from a large-scale model. More complex stochastic weather generators have also been proposed. For example, \cite{kilsby} condition a rainfall model and weather generator on a wet-dry day model, deriving parameters for the model from past and future global climate model runs, thus allowing statistics from the rainfall and weather generators to vary between climate scenarios.

A variety of methods have been developed to account for differences between aggregated and point-level extremes. With the goal of understanding future fine-scale extreme rainfall, \cite{hunting} and \cite{kallache} use similar approaches that establish relationships between extremes of past and future epochs through GEVs fitted to annual maxima of rainfall accumulations generated by RCMs. GEVs fitted to annual maxima of past station data are then transformed accordingly to give quantile-based estimates of future point-level extreme rainfall. A similar approach by \cite{fried} uses quantile regression to relate quantiles of the distribution of rainfall accumulation at a given weather station, conditional on it having rained, to output from a spatially aggregated rainfall model. Alternatively, \cite{smith-downscale} develop a regression relationship between return levels estimated from both large-scale and point-level rainfall data and use this relationship to adjust large-scale return levels to represent point level. By using RCM data for future epochs this approach can also be used to give predictions of future point-level rainfall return levels.

\subsection{Data and notation}

For the remainder of this section the following notation will be used: $X_{F,t}(s)$ and $X_{M,t}(s)$ respectively denote annual maxima of field data and aggregated simulator output for an arbitrary location $s \in R$ and time $t$, $t=1, \ldots, T$. The field data will be assumed to represent point-level in which interest here lies without bias. Being the result of aggregation, such an assumption of unbiasedness cannot be made for the simulator output; consequently we propose to convert the simulator output using a smooth function, denoted $g()$, that will correct for scale difference between the data. In general the optimal form for $g()$ may be unknown and in which case non- or semi-parametric forms may be useful, or parametric forms deemed not to impose unwelcome constraints.

\subsection{Spatial interpolation model formulation} \label{interp-spec}

The model to be used for spatial interpolation is based on the hierarchical spatial model introduced in \S\ref{spat-frame}, and is outlined having assumed that a form for $g()$ has been chosen, which for the application to extreme rainfall is discussed in \S\ref{rain}. For the data layer and given GEV location, scale and shape parameters, $\mu(s)$, $\psi(s)$ and $\xi(s)$ respectively, the GEV in which interest lies is assumed to be shared by annual maxima of the field data, so that \[ \label{fieldgev} \big[X_{F,t}(s)\big]~~\text{is}~~GEV\big(\mu_F(s),\,\psi_F(s),\,\xi_F(s)\big)\] for all $s \in R$ and $t=1,\ldots,T$. Once transformed by $g()$ a related GEV is then assumed for annual maxima of the simulator output: \[ \label{modgev} \big[g\big(X_{M,t}(s)\big)\big]~~\text{is}~~GEV\big(\mu_M(s),\,\psi_M(s),\,\xi_M(s)\big).\] The preceding specification therefore allows the two different sources of data, quantifying the same phenomenon but on different scales, to be modelled jointly. Part of our motivation for this joint modelling comes from \cite{andturk} in which results for the joint distribution of maxima and sums of sequences are derived by combining results from extreme value theory and the central limit theorem.

The joint specification is completed by the spatial process layer. For this a GP is assumed which, considering the GEV's location parameter for illustration, may be given by \[ \label{fieldgaus} \big[\mu_\centerdot(s)\big]~~\text{is}~~GP\big(m_\centerdot(s),\,\sigma_\mu^2 c(\,,\,)\big) \] where $\sigma_\mu^2$ is a variance parameter, $c(\,,\,)$ represents a correlation structure and where $\centerdot$ may be replaced with $F$ or $M$ to represent the separate, respective specifications of the field data and simulator output. Allowing different GP specifications between the two data types consequently allows their differences in scale to be absorbed not only by $g()$ but also by the GP. Similar GPs may also be assumed for $\rho_\centerdot(s)=\log\big(\psi_\centerdot(s)\big)$ and $\xi_\centerdot(s)$.

\section{Model checking} \label{check}

We consider a variety of methods for checking the fit of the latent Gaussian extreme value model described in \S\ref{spat-frame} and \S\ref{interp-spec}.

\subsection{Quantile plots} \label{qplot}

First we asses fit of the proposed model by considering the conditional GEV assumption, given in equation \eqref{data1}, using a modification of the quantile plot. The formulation of \S\ref{mod-est}, in which $\psi(s) = \psi$ and $\xi(s) = \xi$, is again used for illustration, though alterations for when GPs are assumed for other combinations of GEV parameters follow naturally. Modification of a standard quantile plot is required due to the GEV's location parameter being random. 

Suppose that at location $s$ we have observed annual maxima $x_1(s), \ldots, x_n(s),$ with ordered counterparts $x^{(1)}(s) \leq \ldots \leq x^{(n)}(s)$, that are assumed to follow a $GEV\big(\mu(s),\,\psi,\,\xi\big)$ distribution. (These should initially be thought of as annual maxima of field data; quantile plots for the simulator output can be achieved by replacing $x_i(s)$ with $\hat g\big(x_i(s)\big)$ throughout.)

Recall from \S\ref{mod-est} that $\mu_i(s)$, $i=1, \ldots N,$ draws from $[\mu(s)\,|\,x_1, \ldots, x_n]$, can be obtained, and then combined with estimates $\hat \psi$ and $\hat \xi$ of $\psi$ and $\xi$ so that a collection of GEV distributions, $\hat G\big(\,;\mu_i(s)\big)$ and corresponding inverse functions $\hat G^{-1}\big(\,;\mu_i(s)\big)$, $i = 1, \ldots, N$, that reflect the randomness of $\mu(s)$ can be specified. A quantile plot appropriate for the present MCEM setting may then be formed by plotting the pairs \[ \left(x^{(k)}(s),\,\dfrac{1}{N} \sum_{i=1}^N \hat G^{-1}\Big(\frac{k}{n+1};\,\mu_i(s)\Big)\right), \quad k=1, \ldots, n.\label{qq1} \] 

Deviation from linearity of the pairs indicates model failure. The level of deviation expected may be estimated through Monte Carlo simulation, repeatedly sampling from $\hat G\big(\,;\mu_i(s)\big)$. Take $N_G$ samples from $\hat G\big(\,;\mu_i(s)\big)$ and denote the ordered samples by $\hat x_i^{(1)}(s) \leq \ldots \leq \hat x_i^{(n)}(s)$, $i=1, \ldots, N_G$; then take the $l$th order statistic from each sample, ie. $\hat x_1^{(l)}(s), \ldots, \hat x_{N_G}^{(l)}(s)$, and denote their ordered counterparts by $\hat x_{(l)}^{(l)}(s) \leq \ldots \leq x_{(l)}^{(N_G)}(s)$, $l=1, \ldots, n$. Finally, approximate $100(1-\alpha)$\% confidence bounds for the quantile plot at $x^{(i)}(s)$ are given by $\big(\hat x_{(k)}^{(\lfloor N_G\alpha/2\rfloor)}(s),\, \hat x_{(k)}^{(\lfloor N_G(1-\alpha/2)\rfloor)}(s)\big)$ where $\lfloor \cdot \rfloor$ denotes ``integer part''. The accuracy of these confidence intervals can be improved by also accounting for uncertainty in the estimates $\hat \psi$ and $\hat \xi$, and also of $\hat g(\,)$ when using the simulator output; however this modification tends to bring little change to the confidence bounds achieved.

\subsection{Spatial structure diagnostics} \label{spacediag}

This diagnostic is designed to assess the adequacy of the estimated spatial structure of the proposed model by considering how well it compares with empirical estimates of spatial dependence. Again we assume that $[X_t(s) \,\ | \,\mu(s)]$ is $GEV\big(\mu(s), \, \psi, \, \xi\big)$. However, unlike the other model checks, this check does not extend readily to the case in which either or both of $\psi(s)$ and $\xi(s)$ are random, but is sufficient here given the formulation that we adopt when modelling extreme rainfall in \S\ref{rain}. When $[X_t(s)\,|\,\mu(s)]~\text{is}~GEV\big(\mu(s), \, \psi,\, \xi\big)$ we can write \[ X_t(s) = \mu(s) + \varepsilon(s) \] where \[ \big[\varepsilon(s)\big]~\text{is}~GEV\big(0, \, \psi, \, \xi\big) \] and $\mu(s)$ is as in relation \eqref{proc1}. Let $\text{var}\big[\varepsilon(s)\big] = \sigma_\varepsilon^2(s)$. Then for arbitrary $s, s' \in R$ we have that \begin{align*} \text{cov}\big(X_t(s),\, X_t(s')\big) & = \text{cov}\big(\mu(s) + \varepsilon(s),\, \mu(s') + \varepsilon(s')\big)\\ & = \text{cov}\big(\mu(s),\, \mu(s')\big) + \text{cov}\big(\varepsilon(s),\, \varepsilon(s')\big) + \text{cov}\big(\mu(s), \, \varepsilon(s')\big) + \text{cov}\big(\mu(s'), \, \varepsilon(s)\big). \end{align*} The conditional independence of $\varepsilon(s)$ and $\varepsilon(s')$ given  $\mu(s)$ and $\mu(s')$ and independence between $\varepsilon()$ and $\mu()$ gives $\text{cov}\big(X_t(s), \, X_t(s')\big) = \sigma_\mu^2 c(s, s')$ so that \begin{equation} \text{corr}\big(X_t(s),\,X_t(s')\big) = \dfrac{c(s, s')}{\sqrt{\left(1 + \dfrac{\sigma_\varepsilon^2(s)}{\sigma_\mu^2}\right)\left(1 + \dfrac{\sigma_\varepsilon^2(s')}{\sigma_\mu^2}\right)}} \label{eqdiag}\end{equation} Thus a plot of empirical estimates of $\text{corr}\big(X_t(s), \, X_t(s')\big)$ against those expected under the model, given in equation \eqref{eqdiag}, provides a method of assessing the model's spatial structure. Combinations of both field data and simulator output can be assessed by transforming annual maxima by $\hat g()$ where appropriate. Note that for the GEV if $\xi < 0.5$ then $\sigma_\varepsilon^2(s)$ is finite, given by $\psi^2 \{\Gamma(1 - 2 \xi) - \Gamma^2(1 - \xi)\} / \xi^2$ if $\xi \neq 0$ and by $\psi^2 \pi^2 / 6$ if $\xi = 0$. 

In the case where $\psi(s)$ or $\xi(s)$ or both are random, the above procedure cannot be easily modified to provide a similar method of assessing any estimated covariance structure. However simulations from the model may instead be used to provide model-based estimates of $\text{corr}\big(X_t(s), \, X_t(s')\big)$ which may be compared with empirical estimates.

\subsection{Crossvalidation} \label{crossvalid}

A final way in which the fit of the model can be assessed is through crossvalidation, using kriging to predict annual maxima at locations with data though omitted during model estimation. While well documented in the literature, the procedure used is outlined again here as it will be relied on later for interpolation. Let $s^*$ denote a location for which a prediction is required and suppose that ${\bm \mu}_i(s)$, $i = 1, \ldots, N$, have been simulated from $[{\bm \mu}(s)\,|\,{\bf x}]$; then we wish to simulate from $[\mu(s^*)\,|\,{\bm \mu}(s) = {\bm \mu}_i(s), \, {\bf x}]$. This is possible through properties of the Gaussian process as \[ \left[\Big(\begin{array}{c} {\bm \mu}(s)\\ \mu(s^*)\end{array}\Big)\right]~\text{is}~GP\left(\Big(\begin{array}{c} {\bf m}(s)\\ m(s^*)\end{array}\Big),\,\Sigma^*\right)\] where
\[ \Sigma^* = \left(\begin{array}{cc} \sigma_\mu^2c(\,,\,) & \sigma_\mu^2c(\,,s^*)\\ \sigma_\mu^2c(\,,s^*)^T &  \sigma_\mu^2\end{array}\right) = \left(\begin{array}{cc} \Sigma_s & \Sigma_{s,s^*}\\ \Sigma_{s^*,s} &  \sigma_\mu^2 \end{array}\right). \] Then \[ [\mu(s^*)\,|\,{\bm \mu}(s)={\bm \mu}_i(s)]~\text{is}~N\big(\mu_{|s}(s^*),\,\sigma_{|s}^2(s^*)\big) \] where \[ \mu_{|s}(s^*) = m(s^*) - \Sigma_{s^*, s} \Sigma_{s}^{-1} \big({\bm \mu}_i(s)-{\bf m}(s)\big) \] and \[ \sigma_{|s}^2(s^*) = \sigma_\mu^2 - \Sigma_{s^*, s} \Sigma_s^{-1} \Sigma_{s, s^*}.\] 

If GPs are assumed for $\psi(s)$ of $\xi(s)$, kriging may also be used to simulate from their respective conditional distributions; if not the MCEM estimates may be used. The result is that a complete set of GEV parameters may be found for $s^*$ and consequently quantile plots as described in \S\ref{qplot} may be used to assess whether model predictions are consistent with the data not used in model estimation. To account for uncertainty in the kriging estimate due to uncertainty in the parameter estimates on which it depends, simulations from the joint distribution of parameters can be obtained and then kriging estimates produced for each simulation. A potentially more useful application of this kriging-based procedure is the production of return level maps, which will be introduced in the context of extreme rainfall prediction in \S\ref{spatpred}.


\section{Extreme Rainfall} \label{rain}

We now perform spatial interpolation of extreme rainfall using the model introduced in \S\ref{interp}. Attention is restricted to a region that is primarily the South and Midlands of England, indicated in Figure \ref{dataregion}, choosing not to study the entire UK to aid proof of concept of the model. For example, this avoids some of the many coastline effects of extreme rainfall. Extensions to the present analysis, that would help in analysis of the entire UK, are discussed further in \S\ref{rainmodspec}.

\subsection{The data}

\begin{figure}[t]
\begin{center}
\subfloat[\label{dataregion}Study region]{\includegraphics[width=0.25\textheight, trim = 4cm 1cm 4cm 1cm, clip = TRUE]{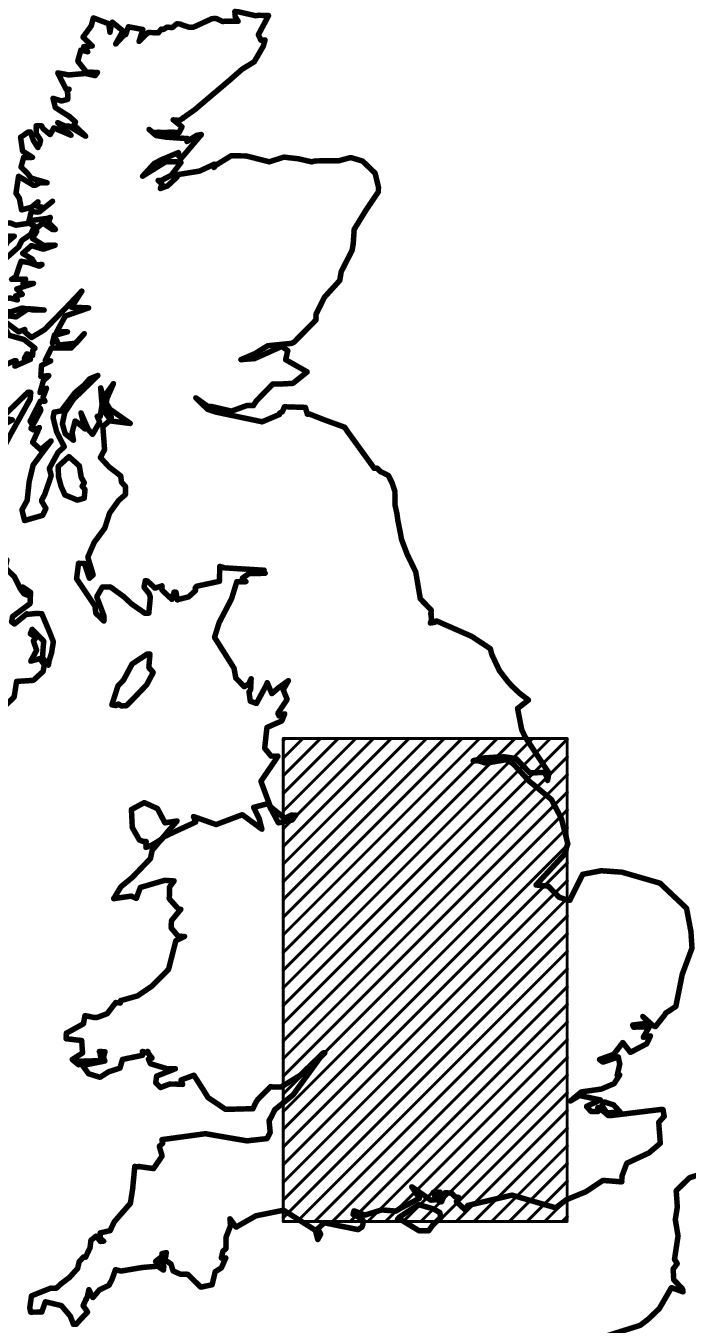}}
\subfloat[\label{dataloc}Data locations by type]{\includegraphics[width=0.25\textheight, trim = 3cm 0cm 3cm 2cm, clip = TRUE]{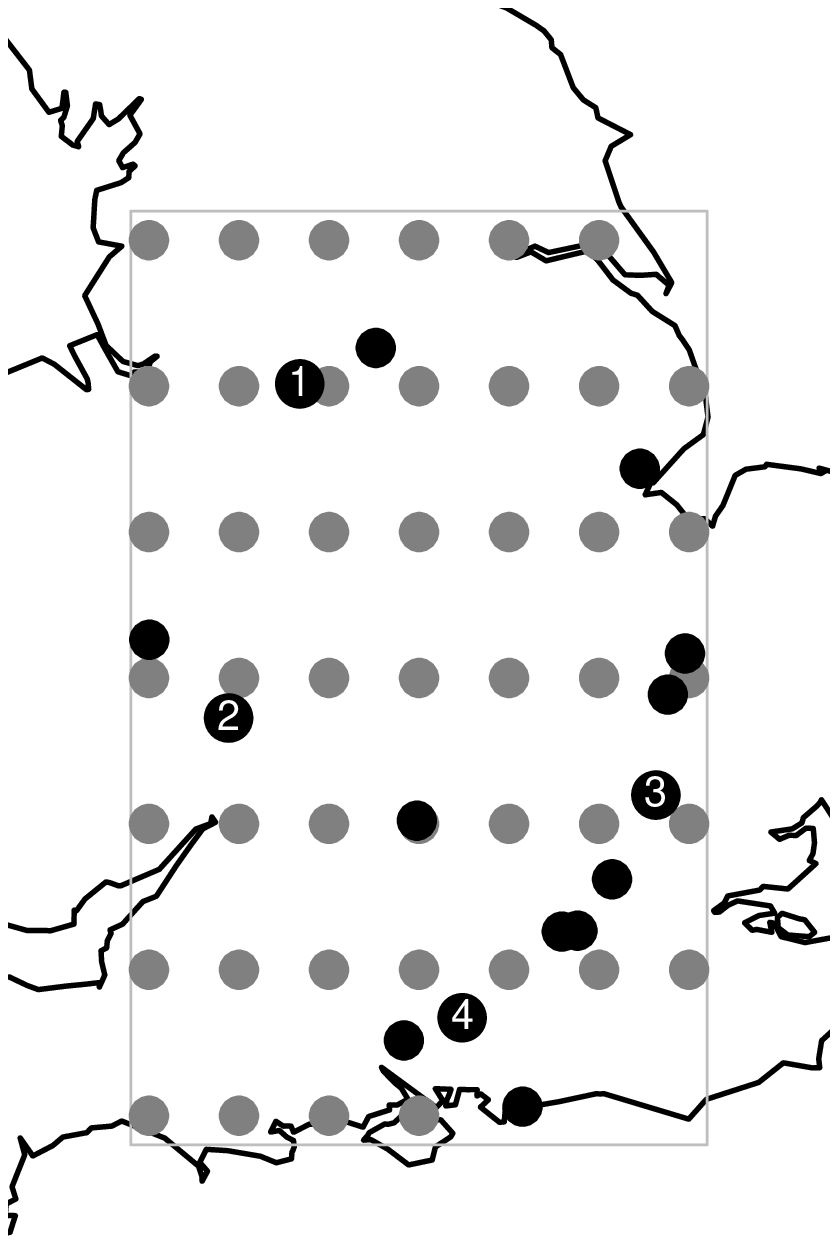}}
\caption{Region of UK studied (left panel) and locations of data used in model estimation, identified by type: (\textcolor{grey}{$\bullet$}) computer simulator output (grid cell mid-point), (\textcolor{black}{$\bullet$}) rain gauge location. (Numbers within symbols identify sites 1, 2, 3, and 4 which are referred to later.)}
\end{center}
\end{figure}

\begin{figure}[p]
\centering
\subfloat[Site 1, distance 10.8km]{\includegraphics[width=0.25\textheight]{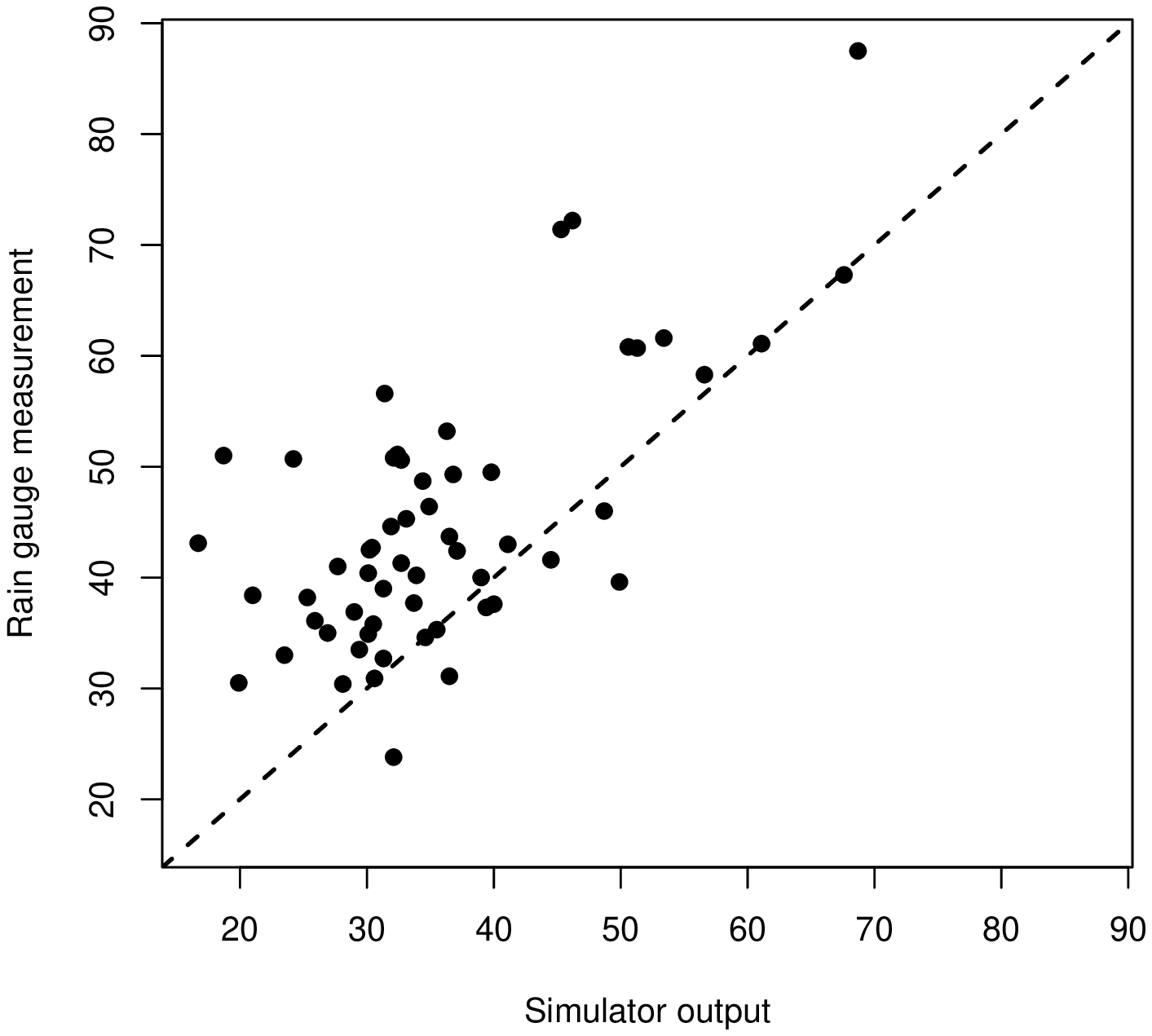}} \qquad
\subfloat[Site 2, distance 15.7km]{\includegraphics[width=0.25\textheight]{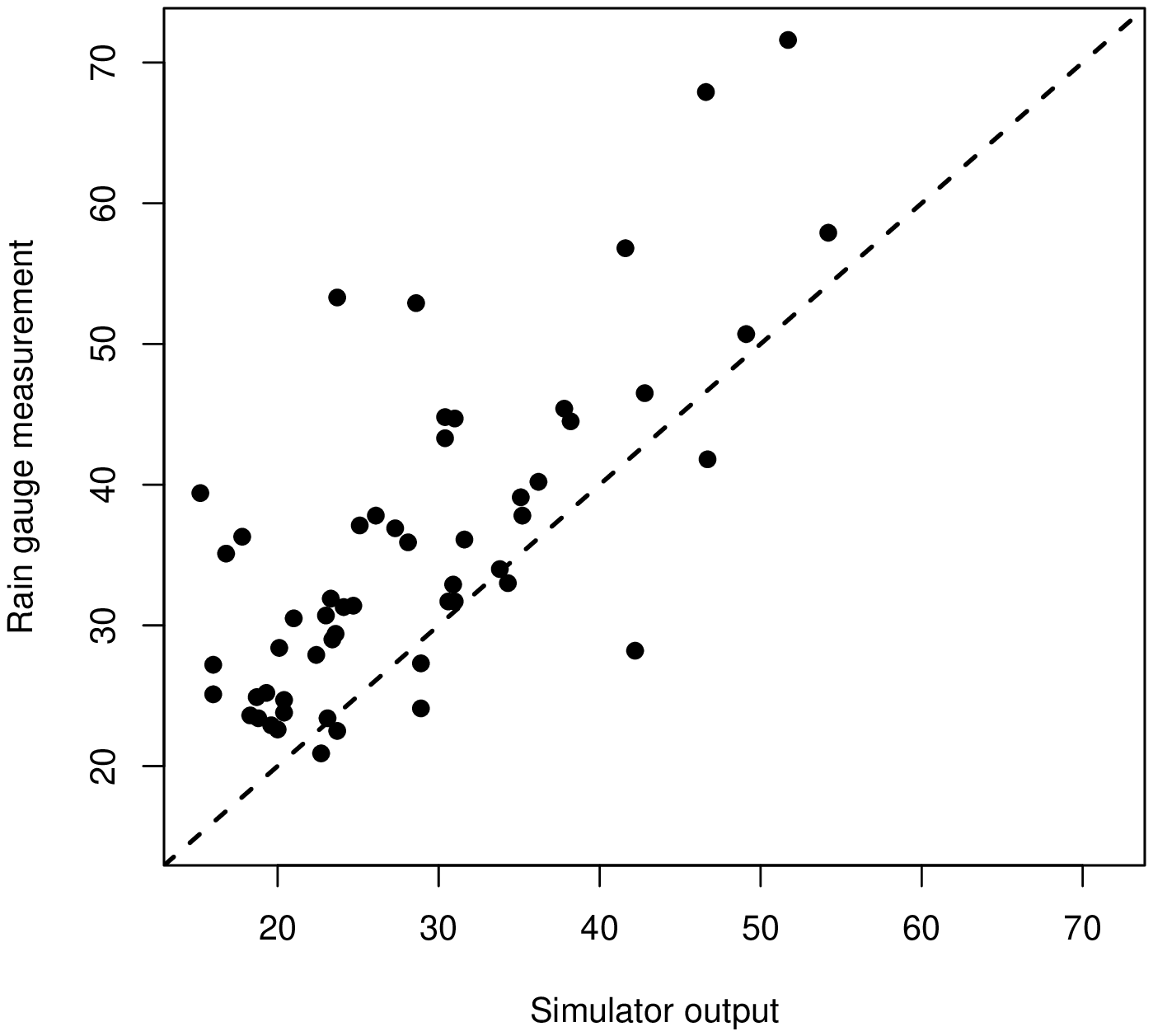}}\\
\subfloat[Site 3, distance 16.8km]{\includegraphics[width=0.25\textheight]{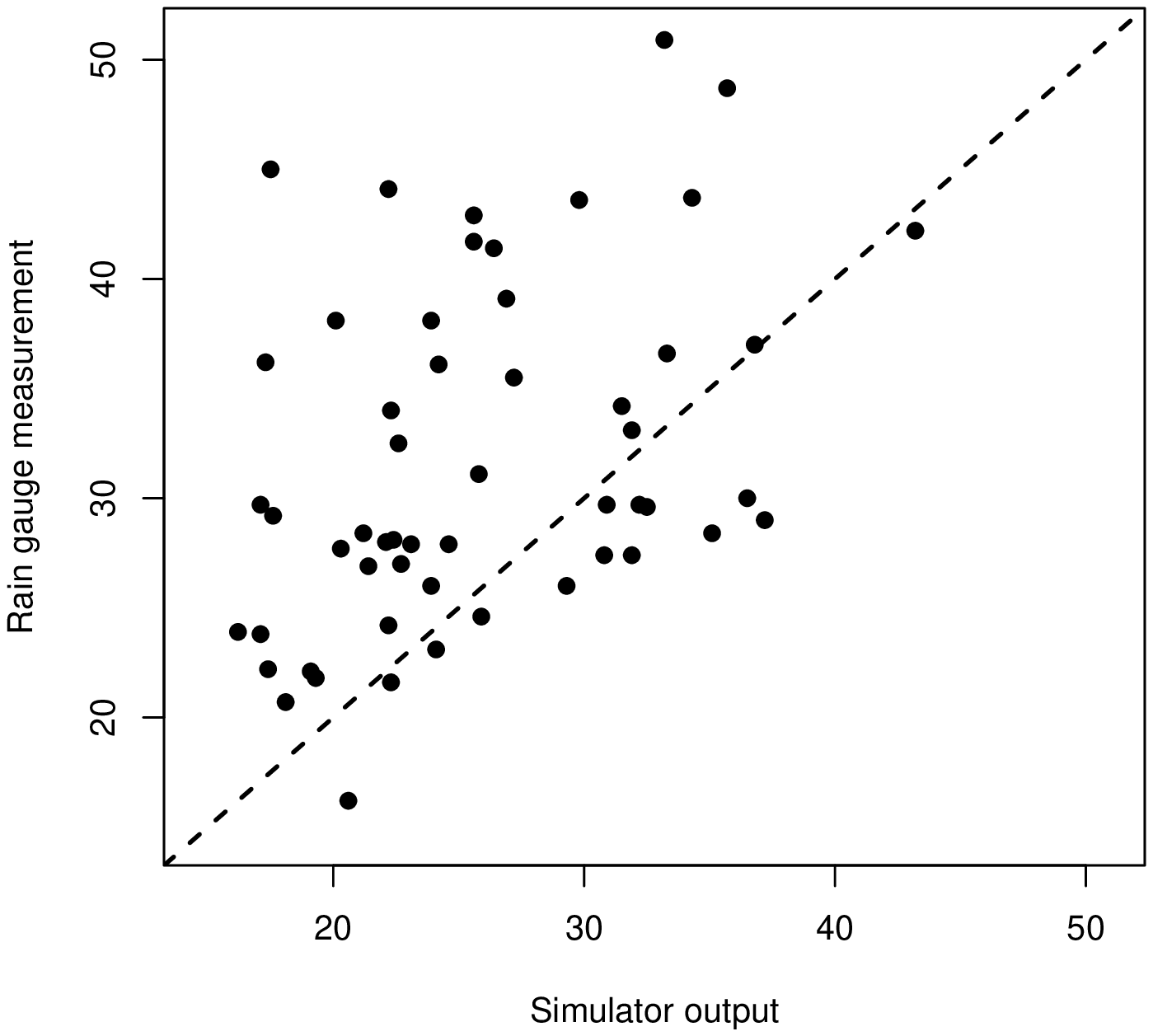}} \qquad
\subfloat[Site 4, distance 24.8km]{\includegraphics[width=0.25\textheight]{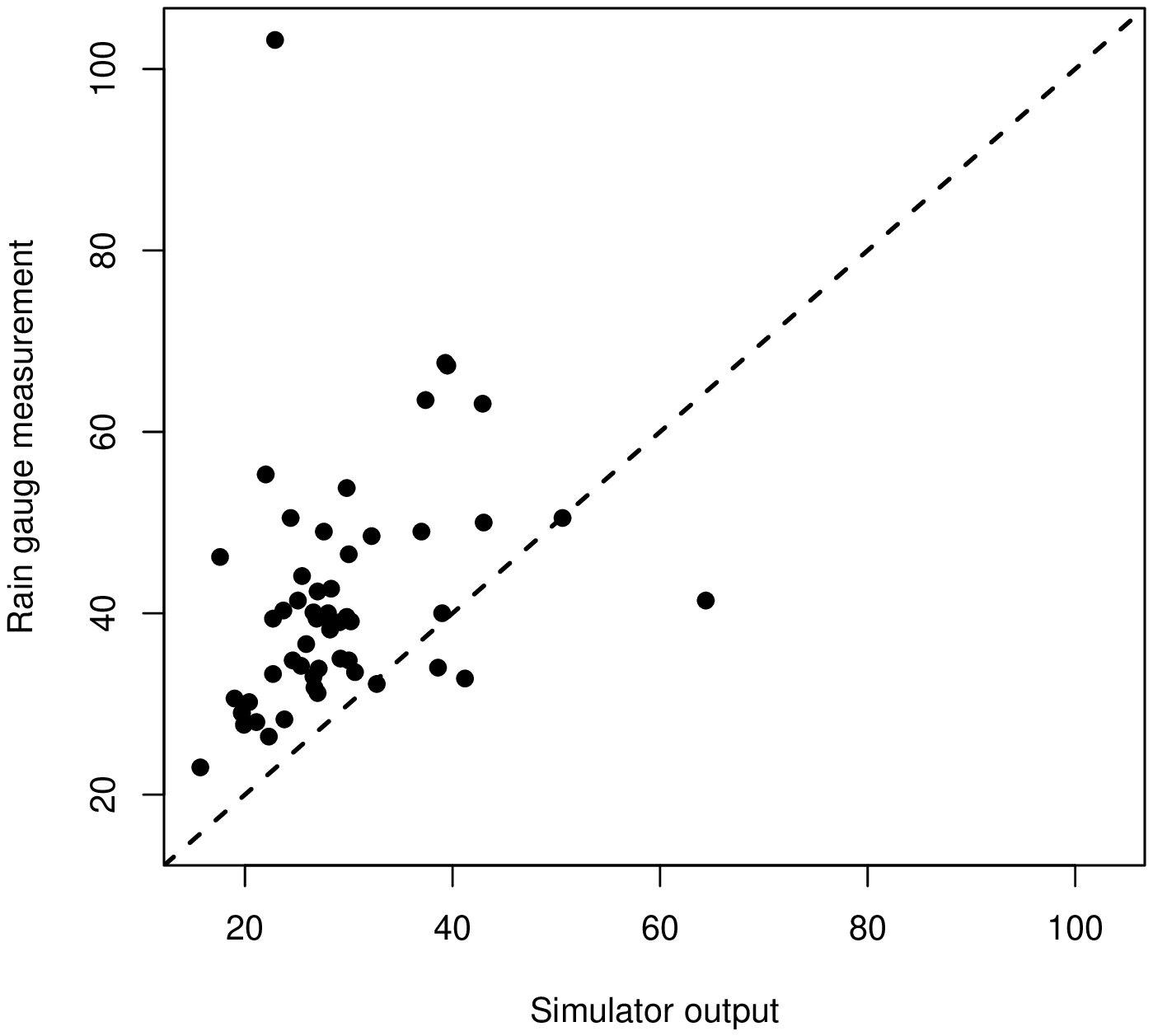}}\\
\caption{\label{compdata}Annual maxima of daily rainfall accumulations (mm) for field data against nearest simulator output. Distance represents that from the location of the rain gauge to the centre of the simulator's corresponding cell. The line of no bias ( - - - ) is superimposed.}
\end{figure}

To estimate model parameters we use both field data and computer simulator output. The field data are annual maxima of daily rainfall accumulations from rain gauges at 15 sites and are obtained from the UK Meteorological Office's MIDAS Land Surface Stations database \citep{ukmetoff2}. The computer simulator output is the $0.5^\circ$ E-OBS gridded dataset \citep{knmidata,knmidata2}, also available as daily data. The locations at which data are used, identified by type, are shown in Figure \ref{dataloc}. Rainfall accumulations from 1st January 1950 to 31st December 2009 are studied. Some years' field data are incomplete, in which case, provided these are believed to be missing at random, annual maxima are omitted from analysis if five or more days' measurements are missing. To give an idea of any systematic differences between the data sources, Figure \ref{compdata} shows plots of field data against most proximate simulator output (defined by distance from rain gauge to nearest grid cell centre) for four locations that are labelled on Figure \ref{dataloc}.

\subsection{Rainfall model specification} \label{rainmodspec}

Particularly important in the model specification is the choice of $g()$, which here we choose first. While the optimal form of the downscaling function is likely to be complex due to the complexity of the computer simulator, a flexible class of model arises from the choice $g(x) = x$, thus absorbing all differences between the different data sources through the GEV parameters and GPs.

\begin{figure}[p]
\begin{center}
\subfloat[Location parameter]{\includegraphics[width=0.9\textwidth]{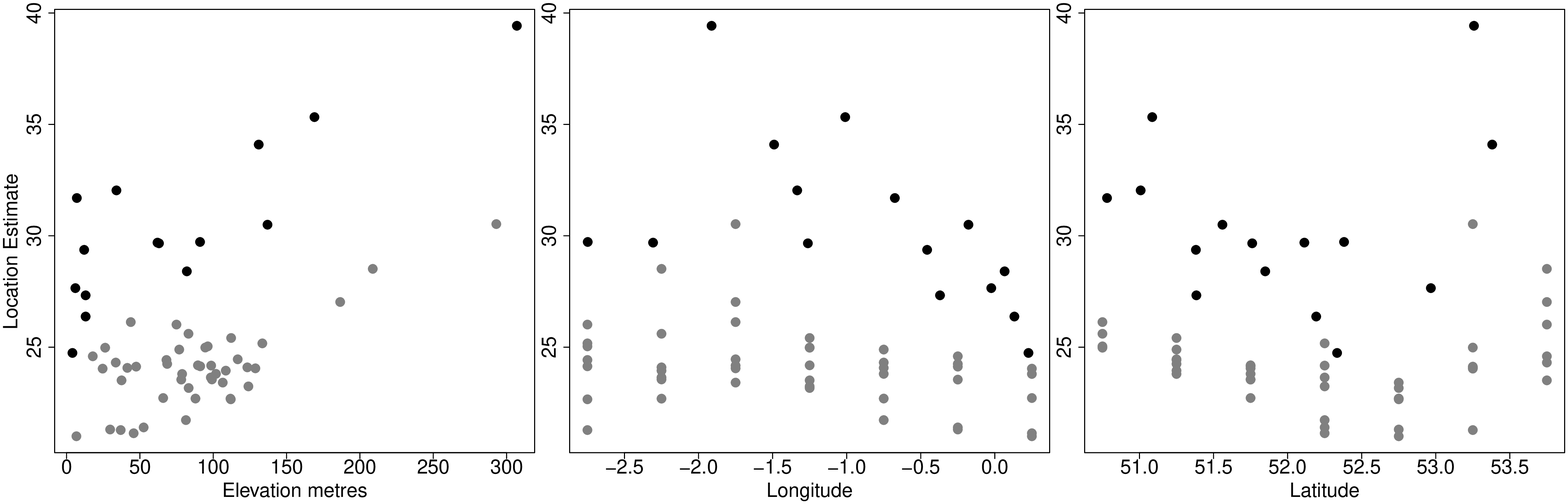}} \\
\subfloat[Scale parameter]{\includegraphics[width=0.9\textwidth]{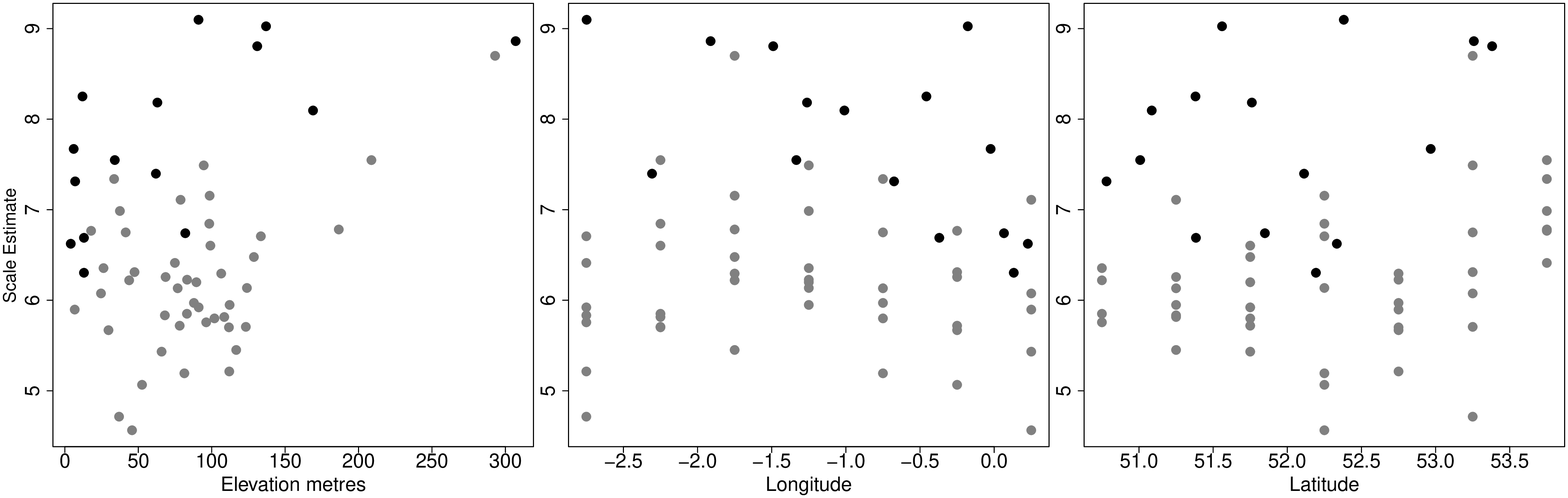}} \\
\subfloat[Shape parameter]{\includegraphics[width=0.9\textwidth]{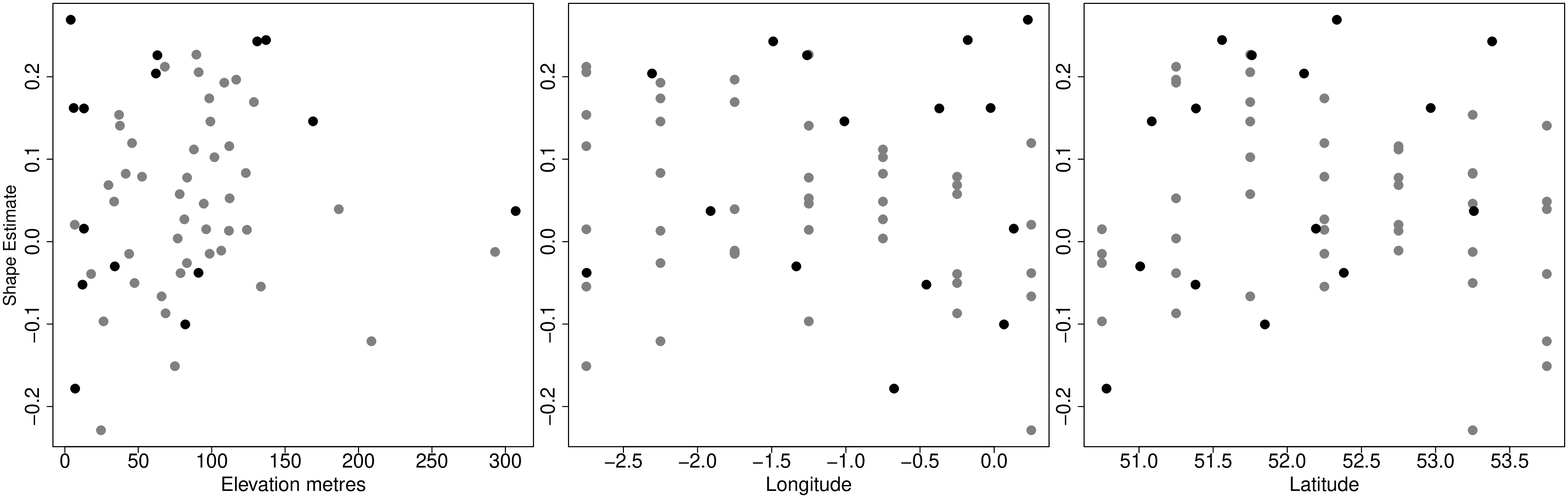}}
\caption{\label{covmarg}Spatially-independent GEV parameter estimates against elevation, longitude and latitude, identified by type: (\textcolor{grey}{$\bullet$}) computer simulator output, ($\bullet$) observational data. }
\end{center}
\end{figure}

A variety of model specifications based on \S\ref{interp-spec} are explored, beginning with assuming GPs for all three GEV parameters, for each of which a variety of mean structures, based on covariates known to influence extreme rainfall, are considered. Initially covariates that may benefit the mean structure are assessed through marginal GEV parameter estimates, that is based on fitting GEVs independently to annual maxima at each location. For each of the GEV's three parameters, plots of parameter estimates against elevation, longitude and latitude are shown in Figure \ref{covmarg}. When considering elevation as a covariate we note that its definition differs between the field data and simulator output: for the former it is simply the height above sea level of the rain gauge, whereas for the latter it represents elevation aggregated over the cell corresponding to the output. These differing definitions suggest using a separate trend in elevation for each data source, which is accommodated through the GP mean structure. Separate trends will also be explored for the longitude and latitude covariates because extreme rainfall quantified by the different data sources could react differently to changes in longitude or latitude, but not because of differing definitions.

Many logical functional forms to capture relationships between the GEV parameters and covariates are studied. These are initially assessed through regression on the marginal parameter estimates, and later through effects of choice of GP mean structures on the MCEM likelihood, specifically the size of the likelihood relative to the number of model parameters. Irrespective of the mean structures for the GEV scale and shape parameters, or whether one or both of the parameters have GP form, their corresponding GP variance estimates are negligibly small. Consequently a GP structure is only adopted for the GEV's location parameter. Models in which $\xi$ is constant, but differs between the data sources, are found to be most parsimonious. Thus any covariate effects are absorbed by the GEV's location and scale parameters. Both parameters are found to depend heavily on elevation, for which different linear trends are assumed between parameters and between data sources. Finally we find the GEV's location parameter to also vary with latitude and longitude, and incorporate this in the model through linear trends that differ between data sources. 

Using the $\centerdot$ notation as in \S\ref{interp-spec}, the final model used is given by \begin{align*} \big[X_{\centerdot,t}(s)\,|\,\mu_\centerdot(s)\big]~~&\text{is}~~GEV\big(\mu_\centerdot(s),\,\psi_\centerdot(s),\,\xi_\centerdot\big)\\ \intertext{where} \big[\mu_\centerdot(s)\big]~~&\text{is}~~GP\big(m_\centerdot(s),\,\sigma_\mu^2 c(\,,\,)\big),\end{align*} with $\sigma_\mu^2$ a variance parameter, $c(\,,\,)$ represents the powered exponential structure described in equation \eqref{powexp} and where \begin{align*} m_\centerdot(s) &=\; \mu_{\centerdot, 0} + \mu_{\centerdot, 1} \times elevation(s)\\& \hspace{1.5cm} + \mu_{\centerdot,2} \times latitude(s) + \mu_{\centerdot,3} \times longitude(s)\\ \intertext{and} \psi_\centerdot(s) &= \exp\{\psi_{\centerdot,0} + \psi_{\centerdot,1} \times elevation(s)\}.\end{align*}

If the preceding analysis was to be extended to modelling extreme rainfall for the entire UK, one of the most significant changes that might benefit the above model would be to consider proximity of locations to the coast, and consequently to also possibly account for the direction of prevailing winds, and to incorporate these through further covariates. 

%
%
%

\subsection{Model estimates}

\begin{figure}[t]
\centering
\subfloat[$\psi_{F,0}$]{\includegraphics[width=0.45\textwidth]{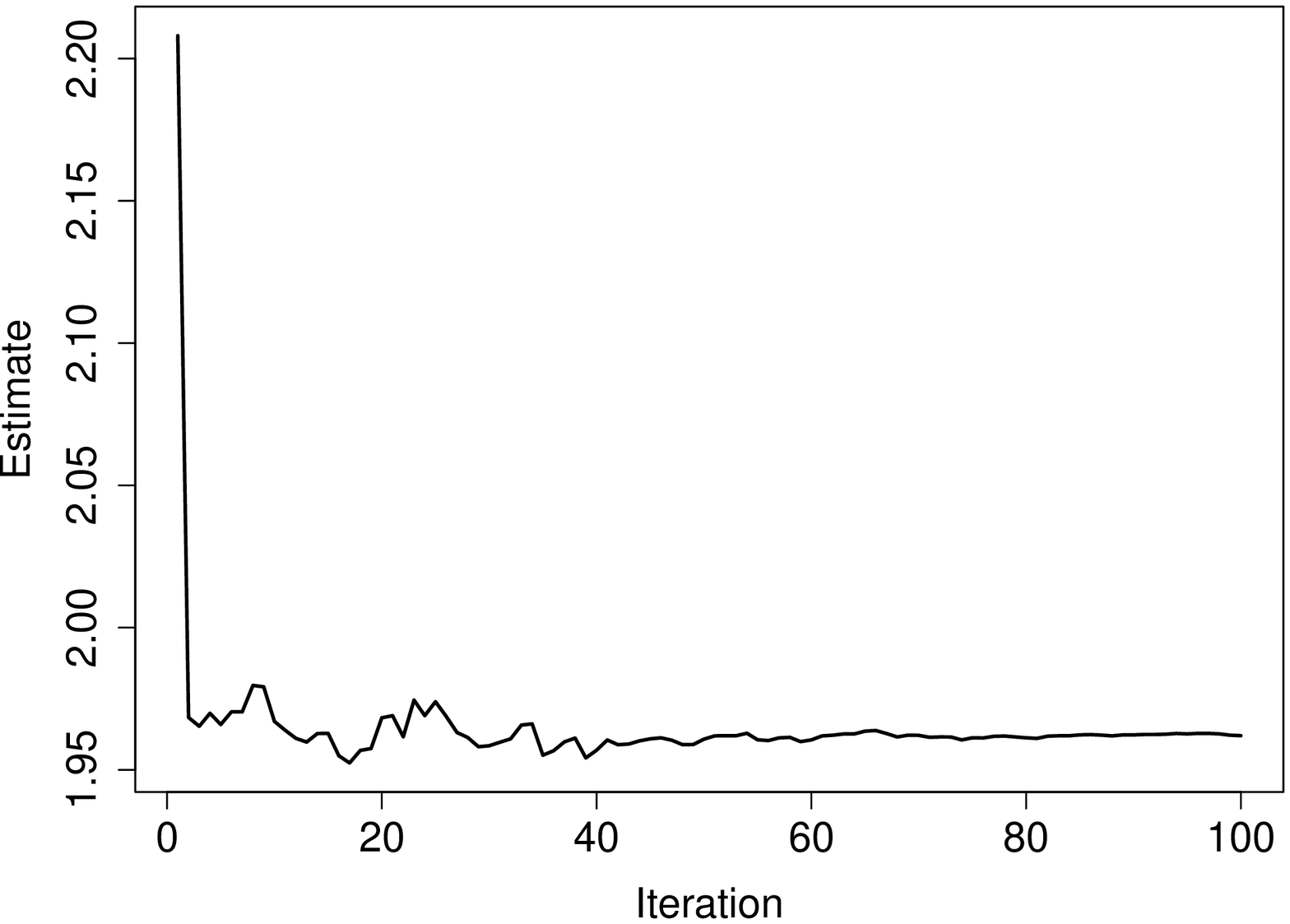}}  \qquad
\subfloat[$\psi_{M,0}$]{\includegraphics[width=0.45\textwidth]{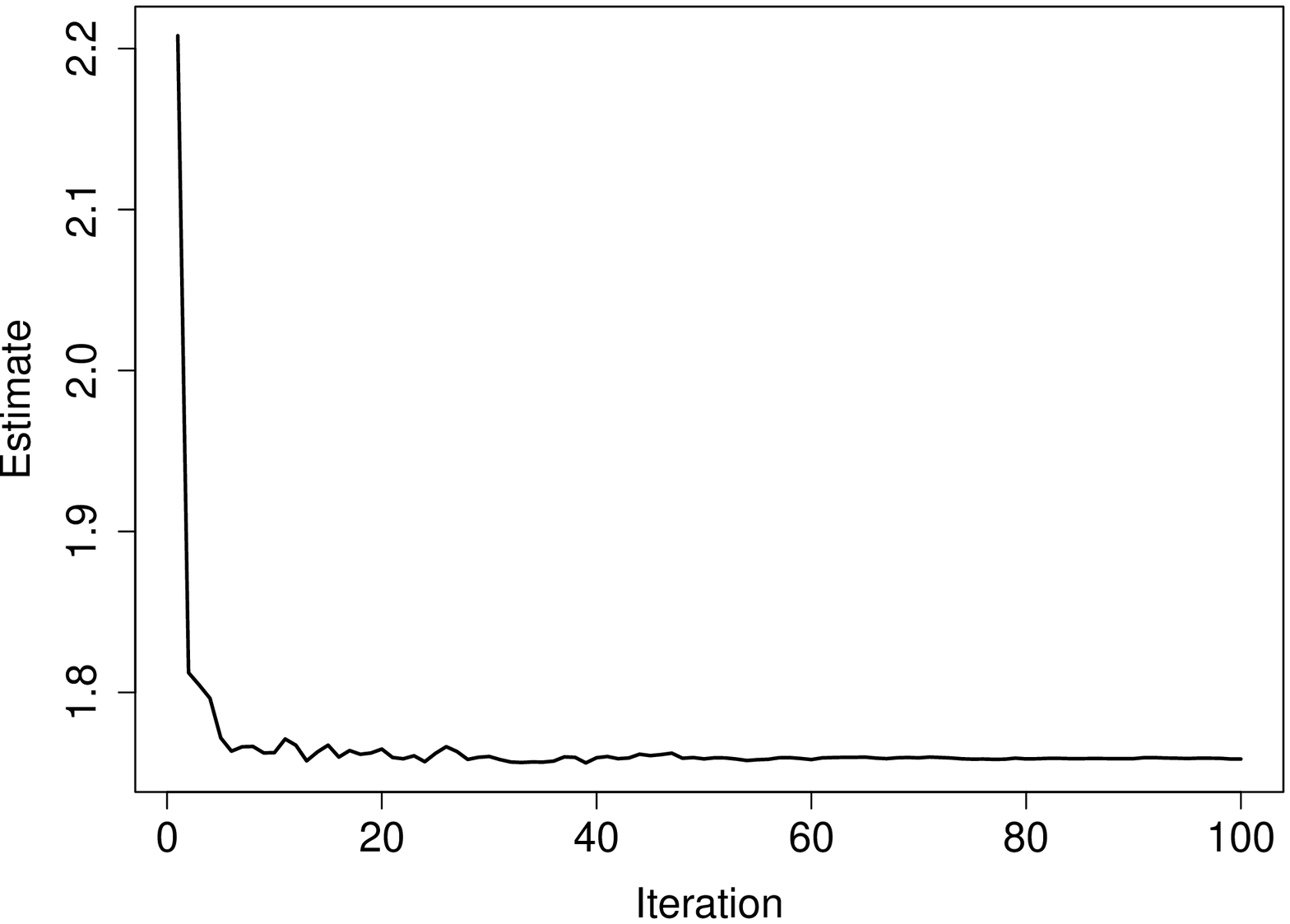}}\\
\subfloat[$\mu_{F,0}$]{\includegraphics[width=0.45\textwidth]{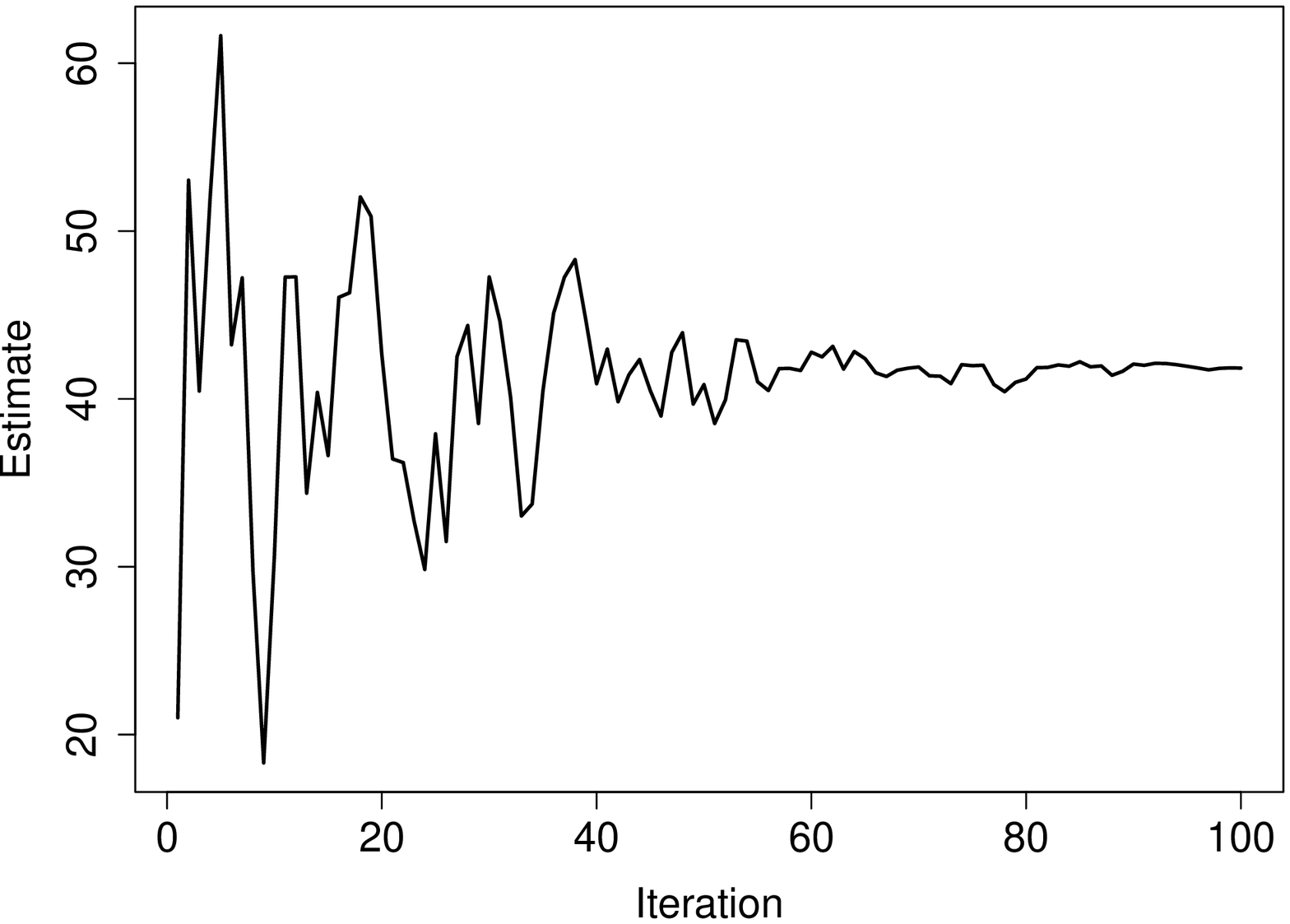}}  \qquad
\subfloat[$\mu_{M,0}$]{\includegraphics[width=0.45\textwidth]{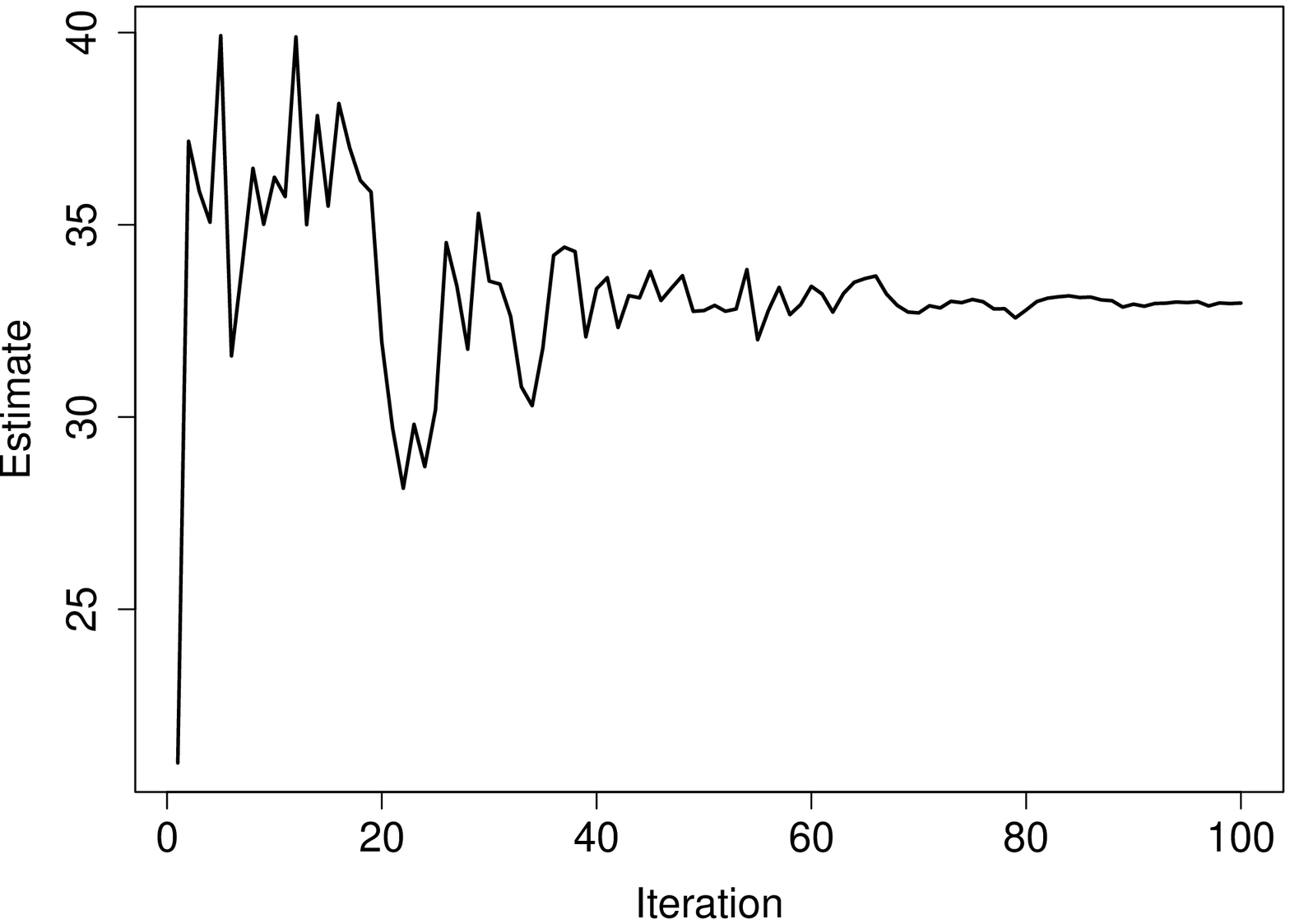}}\\
\caption{\label{parconv}Parameter estimates for $\psi_{\centerdot,0}$  and $\mu_{\centerdot,0}$ at each of the 100 iterations of the MCEM algorithm. Initially $N=10D$.}
\end{figure}

All of the parameters estimated were introduced in \S\ref{rainmodspec}. Estimates of $\psi_{F,1}$ and $\psi_{M,1}$ from the data layer of the model, and of $\mu_{F,0}$ and $\mu_{M,0}$ from the spatial process layer, are shown for each iteration of the MCEM algorithm in Figure \ref{parconv}; convergence appears convincing and as a result the MCEM method of parameter estimation is deemed to work well. Note that altogether we have data for $D = 60$ sites and perform 100 iterations. Initially for the MCEM algorithm we choose $N = 10D$ and increase this by 10\% at each iteration. By gradually increasing $N$ to its final value the speed of convergence is improved because an approximate estimate is reached quickly and is then made more accurate by the increase in $N$. This procedure also helps avoid finding only local as opposed to global maxima. Alternative initial parameter values were also tested, though all led to the same final estimates. Table \ref{parest} \begin{table}[t]
\begin{center}
\begin{tabular}{cll||cr@{.}lr@{.}l}
\multicolumn{3}{c||}{Data layer} & \multicolumn{4}{c}{Spatial process layer}\\
	Parameter & Estimate & S.E. & Parameter & \multicolumn{2}{l}{Estimate} & \multicolumn{2}{l}{S.E.}\\ \hline
$\psi_{F,0}$ & 1.96 & 0.0659 & $\mu_{F, 0}$ & 41&8 & 14 & 7\\
$\psi_{F,1}$ & 0.000782 & 0.000610 & $\mu_{F, 1}$ & 0&0342 & 0 & 00206\\
$\psi_{M,0}$ &  1.76 & 0.0180 & $\mu_{F, 2}$ & -0&371 & 0 & 228\\
$\psi_{M,1}$ & 0.000986 & 0.000191 & $\mu_{F, 3}$ & -0&276 & 0 & 283\\
$\xi_F$ & 0.101 & 0.0642 & $\mu_{M, 0}$ & 33&0 & 10 & 3\\
$\xi_M$ & 0.050 & 0.00766 & $\mu_{M, 1}$ & 0&0223 & 0 & 00201\\
& & & $\mu_{M, 2}$ & -0&162 & 0 & 167\\
& & & $\mu_{M, 3}$ & -0&205 & 0 & 197\\
& & & $\sigma_\mu$ & 0&0121mm & 0 & 131\\
& & & $\phi$ & 3&84km & 0 & 845\\
& & & $\delta$ & -0&643 & 0 & 271\\
& & & $\tau$ & 0&050km & \multicolumn{2}{c}{N/A}\\
\end{tabular}
\vspace{0.5cm}
\caption{\label{parest}Parameter estimates for the model described in \S\ref{rainmodspec}. Note that $\tau$ is fixed, and consequently has no S.E. estimate.}
\end{center}
\end{table}shows estimates for all parameters based on iteration 100. The accompanying standard error estimates for the data layer are achieved using the variant of the sandwich estimator introduced in \S\ref{sandwich}, whereas those for the spatial process layer are based on the usual observed Fisher information. 

\subsection{Model checks}

\begin{figure}[p]
\centering
\subfloat[Site 1]{\includegraphics[width=0.25\textheight]{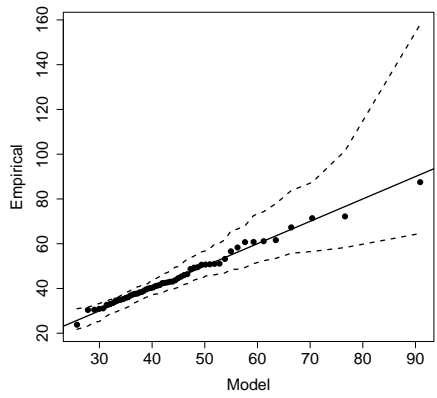}} \qquad
\subfloat[Site 2]{\includegraphics[width=0.25\textheight]{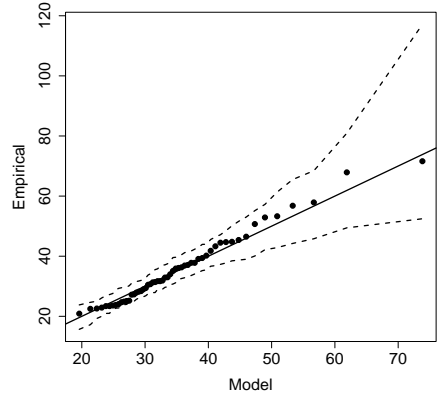}}\\
\subfloat[Site 3]{\includegraphics[width=0.25\textheight]{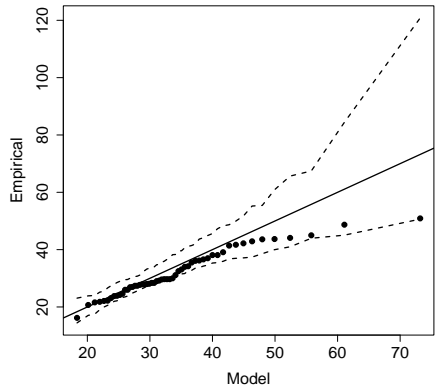}} \qquad
\subfloat[Site 4]{\includegraphics[width=0.25\textheight]{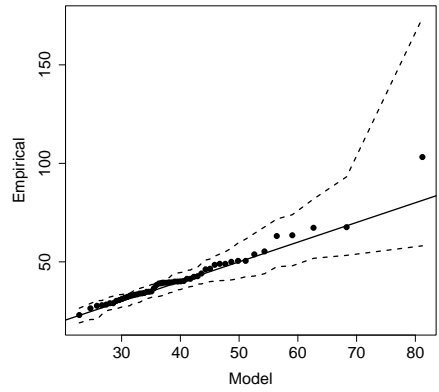}}\\
\caption{\label{quantplot1}GEV quantile plots of annual maxima from the field data against those expected under the model together with 95\% confidence bounds for sites identified in Figure \ref{dataloc}.}
\end{figure}

Initially the fit of the model is assessed using quantile plots, outlined in \S\ref{qplot}. These are shown in Figure \ref{quantplot1}. The plots for almost all sites do not give reason to doubt the estimated model, as the points deviate little from linearity. For site 3, for example, this deviation is larger than for the other sites, and its form suggests that the annual maxima at that site may be consistent with a GEV with a lighter tail. However, as this deviation is within the confidence bounds given, and because in general the field data appear consistent with the estimated spatial model, the present check does not give cause for concern. Furthermore, while not shown in the present paper, related quantile plots for the simulator output, using the method mentioned in \S\ref{qplot}, are equally supportive of the estimated model.

\begin{figure}[t]
\centering
\includegraphics[width=0.5\textwidth]{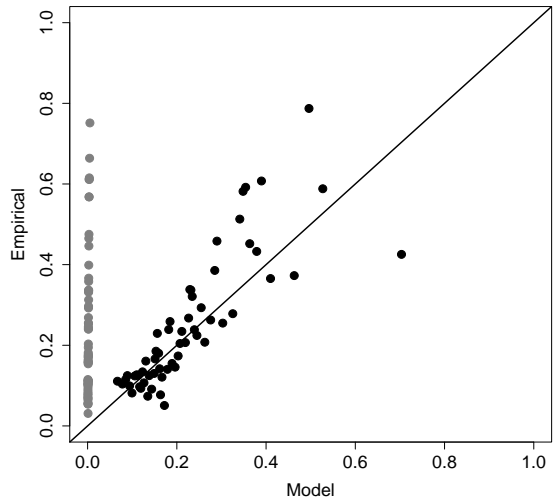}
\caption{\label{spaceplot}Plots of binned empirical against model correlation estimates for $k = 0$ (\textbullet) and $k = 1$ (\textcolor{grey}{\textbullet}).}
\end{figure}

We proceed by using the method outlined in \S\ref{spacediag} to assess the fit of the estimated spatial structure. Upon simple inspection there are signs that conditional on the random GEV location parameters, the remaining variability in annual maxima is notably less than that of the assumed GEV distribution. Consequently we modify the estimate of equation \eqref{eqdiag} so that we simply assume that $\sigma_{\varepsilon,\centerdot}^2(s) = k \hat\psi_\centerdot^2(s) \{\Gamma(1 - 2 \hat \xi_\centerdot) - \Gamma^2(1 - \hat \xi_\centerdot)\} / \hat \xi^2$, noting that $0 < \hat \xi_\centerdot \pm 2 \times \text{S.E.}(\hat \xi\centerdot)< 0.5$, thus assuming that the residual variability is proportional to that expected under the model. Therefore, considering the correlation between annual maxima of the field data and simulator output for example, \begin{equation} \text{corr}\big(X_{F,t}(s), X_{M,t}(s')\big)  = \dfrac{c(s, s')}{\sqrt{\left(1 + \dfrac{\sigma_{\varepsilon,F}^2(s)}{\sigma_\mu^2}\right)\left(1 + \dfrac{\sigma_{\varepsilon, M}^2(s')}{\sigma_\mu^2}\right)}} \label{eqdiag2} \end{equation} for $t=1, \ldots, T$ and $s, \,s' \in R$, noting that $g(x) = x$. Figure \ref{spaceplot} shows a plot of corr$\big(X_{\centerdot,t}(s), X_{\centerdot,t}(s')\big)$ against the estimate in the RHS of equation \eqref{eqdiag2} considering all combinations of field data and simulator output locations. Correlation estimates are binned based on the model-based estimates to ease comparison. The resulting plots of Figure \ref{spaceplot} in general show that the model's estimated spatial dependence structure is consistent with its empirical counterparts once residual variability in annual maxima given respective GEV parameters has been eliminated, ie. when $k = 0$. Without altering the residual variability, ie. taking $k = 1$, we see from Figure \ref{spaceplot} that the empirical correlations between annual maxima are significantly greater than expected under the model, indicating that the original conditional independence assumption, introduced in \S\ref{spat-frame}, is violated. Thus use of the information sandwich correction to estimate standard errors associated with parameters in the data layer is vital for giving adequate estimates of parameter uncertainty.

\subsection{Spatial prediction} \label{spatpred}

\begin{figure}[t]
\centering
\subfloat[100-year return level estimate]{\includegraphics[width=0.35\textheight,trim = 5cm 1.5cm 0cm 1.5cm, clip = TRUE]{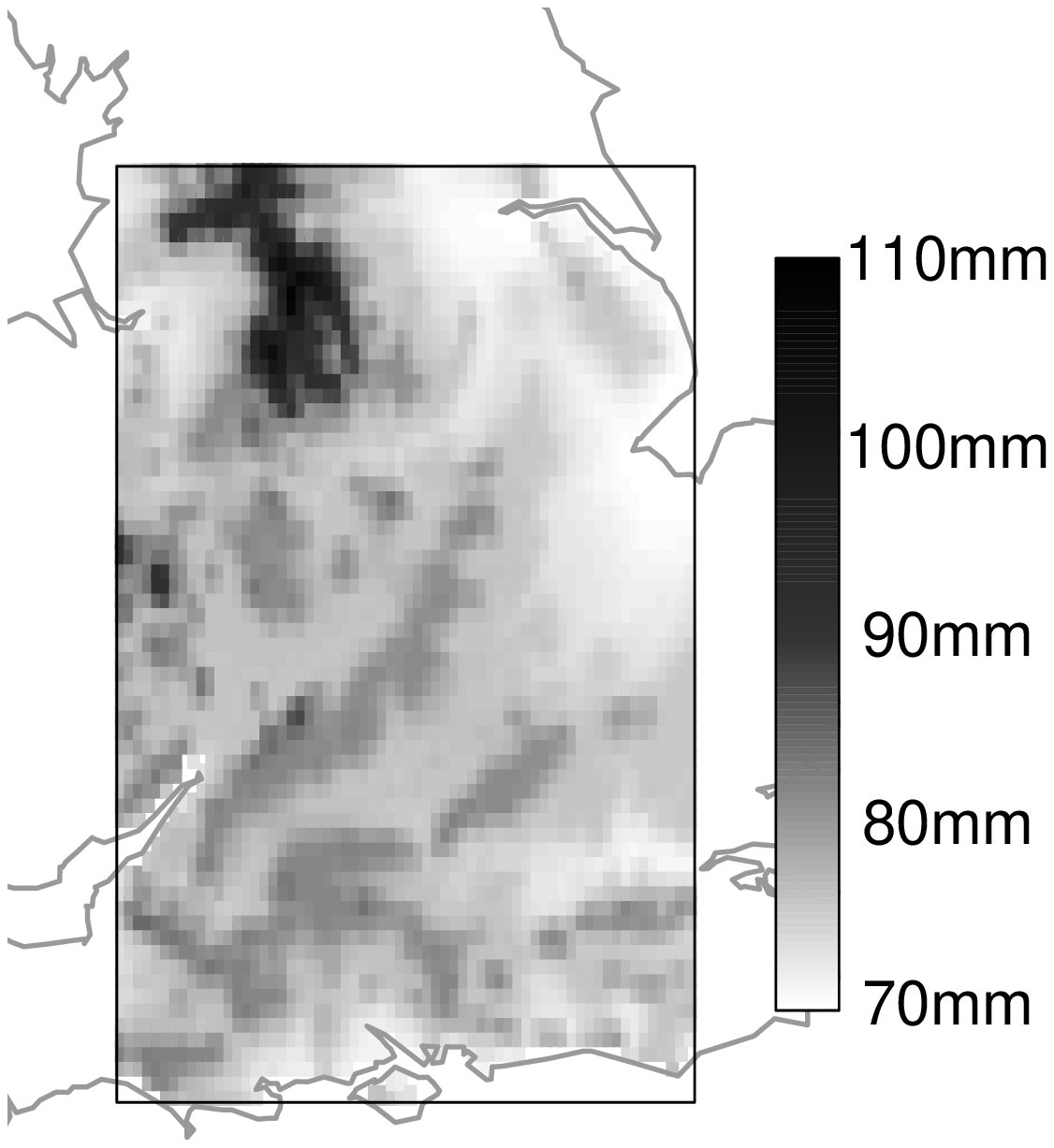}}
\subfloat[95\% confidence interval width]{\includegraphics[width=0.35\textheight,trim = 5cm 1.5cm 0cm 1.5cm, clip = TRUE]{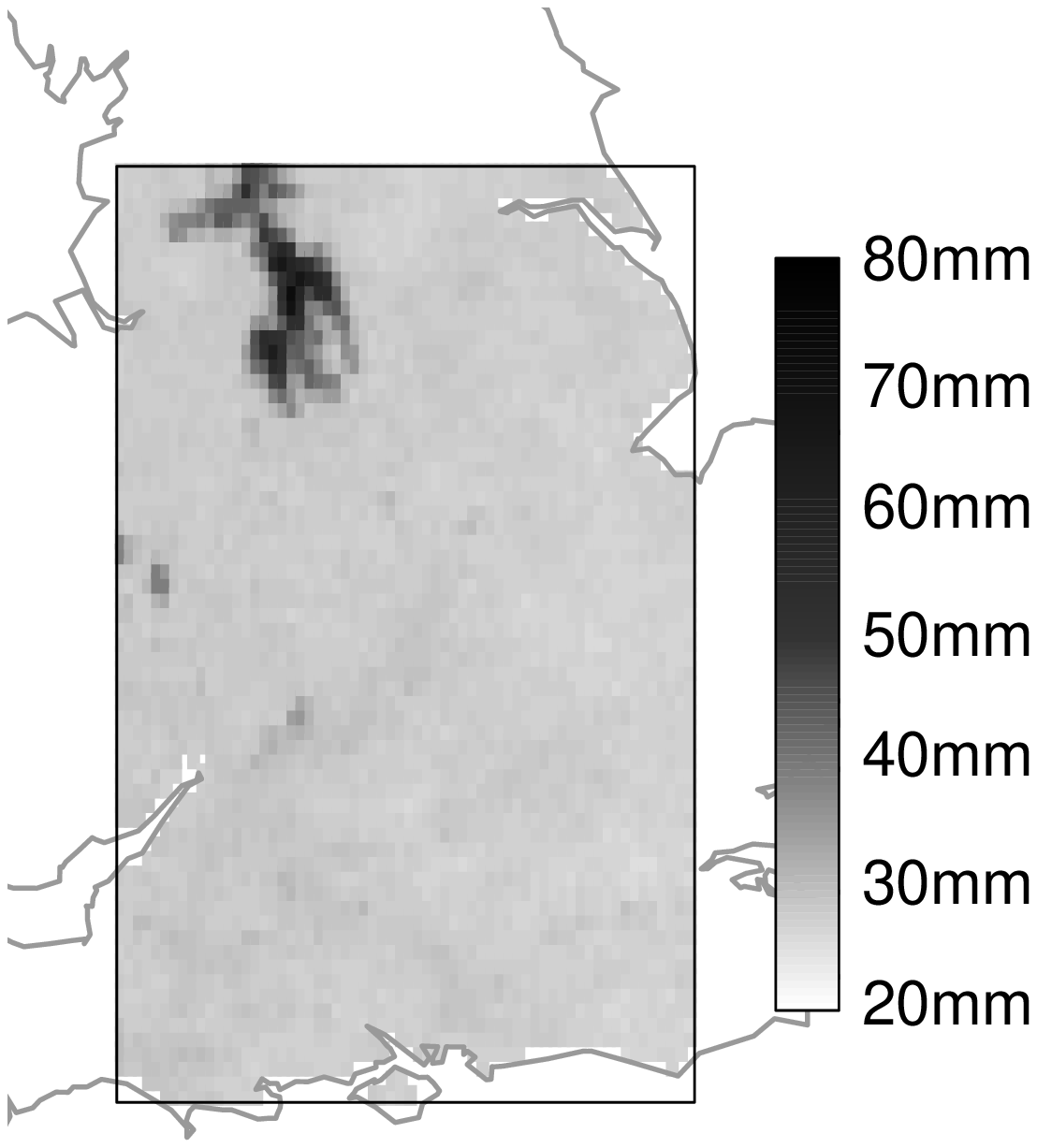}}\\
\caption{\label{rlplot}Return level maps.}
\end{figure}

Finally Figure \ref{rlplot} shows a map of the 100-year return level estimate, together with 95\% confidence bound widths, for the region of the UK under study. The map is obtained from estimates of the 0.99 quantile of the GEV distribution for each location in the region. The multivariate normal distribution from which to simulate GEV scale and shape parameters, and consequently represent their uncertainty accurately, is given by arguments in \S\ref{sandwich}, and uncertainty in the kriging estimate for the GEV location parameter is achieved by the method described in \S\ref{crossvalid}. The location, scale and shape parameter samples can then be combined and to give a return level sample and then variability in the samples used to accurately quantify uncertainty in the return level map.

One of the most prominent features of Figure \ref{rlplot} is its resemblance to a relief map of the region under study. This is a consequence of elevation being the most influential covariate included in the model, which can be seen from its corresponding estimates given in Table \ref{parest}. A further way in which the model's performance can be assessed is by crossvalidation; that is predicting annual maxima at sites with data but deliberately omitted from model estimation. Quantile plots similar to those shown in Figure \ref{quantplot1} can then be used to assess fit. In general these display similar features to those of Figure \ref{quantplot1}, and as a result are not shown, but offer further support for the fit of the present model. Consequently the return level map is deemed to provide a plausible representation of point-level behaviour of the 100-year return level for annual maxima of daily rainfall accumulations.

\section{Discussion} \label{discuss}

In this paper we have provided a method for interpolating extreme rainfall at fine scale based on a coherent way of spatially pooling related though inherently different data. Point-level estimates of extreme rainfall can then be produced for an entire spatial region, which has been achieved here using rain gauge measurements at only a few locations. This estimation would otherwise not be possible if a marginal approach, in which GEVs are fitted independently at different locations, had been used. Furthermore this method offers the potential for estimates of areal rainfall, such as extreme rainfall accumulations for a river catchment area, to be obtained. While we have used measurements from only a few rain gauges, the model is equally applicable if measurements from considerably more gauges were used.

This work has also shown that the MCEM algorithm can be used reliably to provide estimates of parameters in latent Gaussian spatial models for extremes, and introduced a simple diagnostic tool that allows model-based estimates of spatial dependence between annual maxima to be compared with empirical counterparts for the model formulation adopted here. Furthermore we have been able to overcome potential misspecification in the model, in particular violation of the conditional independence assumption, and still give adequate estimates of parameter uncertainty by introducing a variant of the information sandwich estimator applicable to the MCEM algorithm.

\section{Acknowledgements}

I thank C. W. Anderson for many useful discussions that have brought considerable improvement to this work. I also thank the EPSRC for financial support from a Doctoral Training grant and acknowledge the E-OBS dataset from the EU-FP6 project ENSEMBLES (\url{http://ensembles-eu.metoffice.com}) and the data providers in the ECA\&D project (\url{http://eca.knmi.nl}).

\bibliographystyle{chicago}
\bibliography{SpatInterp}

\end{document}